\documentclass[12pt,preprint]{aastex}

\usepackage{psfig,times,emulateapj5}
\usepackage{natbib}
\usepackage[below]{placeins}

\newcommand{\rxj}{RX~J0720.4-3125}

\newcommand{\xspec}{{\em XSPEC}}

\newcommand{\chandra}{{\em Chandra}}

\newcommand\xmm{{\em XMM-Newton}}

\newcommand{\nh}{{$N_{\rm H}$}}

\begin{document}

\title{The continued spectral evolution of the neutron star {RX~J0720.4-3125}}
\author{Jacco Vink, Cor P. de Vries,  Mariano M\'{e}ndez}

\affil{SRON National Institute for Space Research, 
  Sorbonnelaan 2, NL-3584 CA, Utrecht, The Netherlands}
\email{j.vink@sron.nl}
\and
\author{Frank Verbunt}
\affil{Astronomical Institute, Utrecht University, 
   PO Box 80000, NL-3508 TA Utrecht, The Netherlands}

\begin{abstract}
We observed the isolated neutron star \rxj\ with
\chandra's Low Energy Transmission Grating Spectrometer, 
following the \xmm\ discovery
of long term spectral evolution of this source.
The new observation shows that the spectrum of \rxj\
has continued to change in the course of 5 months. 
It has remained hard, similar
to the last \xmm\ observation, but the strong depression observed with \xmm\
at long wavelengths has disappeared. Contrary to the \xmm\ observations,
the new \chandra\ observation shows that the flux increase at short
wavelengths and the decrease at long wavelengths do not necessarily occur
simultaneously.
\end{abstract}


\maketitle
\section{Introduction}
\rxj\ \citep{haberl97}, 
is one of the best studied isolated
neutron stars whose X-ray emission is
dominated by radiation from the hot neutron star surface.
Their X-ray spectra are best described by blackbody emission with
$kT_\infty \sim 70$~eV, but the emitting areas seem
to be significantly smaller than the canonical neutron star surface.
This suggests that the X-ray 
emission is coming from a small, hot, fraction
of the neutron star \citep[][]{motch03,kaplan03b}, 
or may not be pure blackbody
radiation, but instead the result of
emission from a reflective condensed matter surface
\citep[][]{lenzen78,turolla04}.
Recently it was observed that the isolated neutron stars
RX J1308.6+2127 and RX J1605.3+3249 
exhibit broad absorption features that may be caused by
proton cyclotron absorption \citep{haberl03a,vankerkwijk04}.

The spectral behavior of \rxj\ is even more surprising.
Early observations with \chandra\ \citep{kaplan03b} 
and \xmm\ \citep{paerels01} are consistent with a blackbody-like spectrum, 
but subsequent observations by \xmm\ indicate a slow evolution
of the spectrum between 10-38~\AA, 
resulting in deviations from a Planckian spectrum.
The deviations consist of a flux decrease at wavelengths longer
than $\sim 23$\AA,
and a flux increase at shorter wavelengths
\citep{devries04}.
The nature of the emerging spectrum is not clear (modified blackbody,
cyclotron absorption in combination with a hotter blackbody?),
nor is it clear what causes
the emission properties to change. In \citet{devries04} 
we suggested that \rxj\ is precessing,
which causes us to observe the hot region of the star under a 
continuously different angle.
This requires that the surface emission is anisotropic,
and  may alter as result of traversing a magnetized atmosphere
\citep{ho04}.

Whatever the underlying mechanism, 
it is clear that the spectral evolution of \rxj\
potentially provides clues to the nature of the X-ray surface emission
from isolated neutron stars, whereas the fact that the spectrum evolves
may have implications for the structure of neutron stars, e.g.
if the evolution is a consequence of precession.
In order to further monitor its spectral evolution,
and to investigate the spectrum for $\lambda > 38$~\AA\ with high 
spectral resolution,
we were granted Director's Discretionary Time
(DDT) on \chandra\ for observing
\rxj\ with the Low Energy Transmission Grating 
Spectrometer (LETGS) in combination with the HRC-S (High Resolution
Camera) microchannel plate detector.
In this letter we report on the analysis of this observation.
We show that the spectrum has continued to evolve: it has further hardened,
but the attenuation at long wavelengths has disappeared.

\begin{figure*}
\centerline{
\psfig{figure=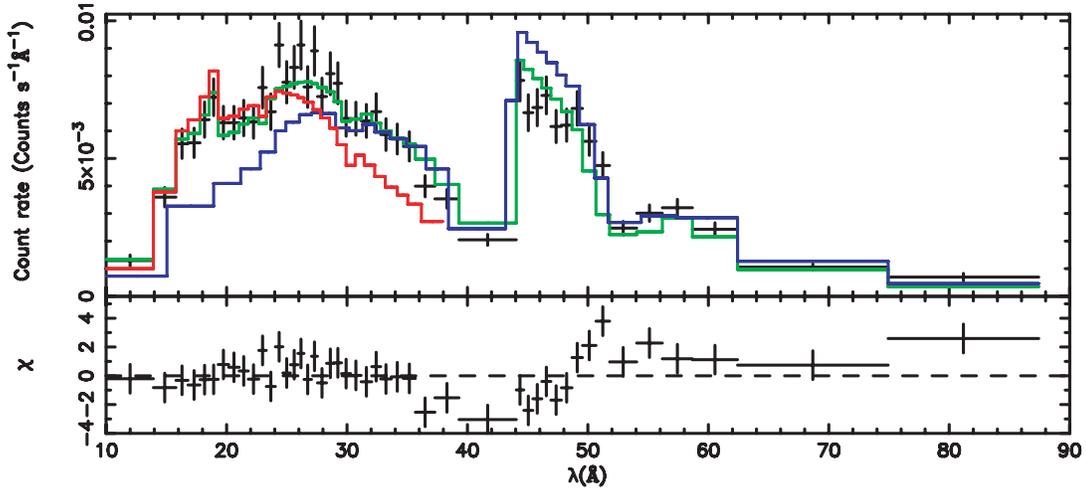,width=0.77\textwidth}}
\figcaption{
\chandra\ LETGS spectrum of \rxj\ as observed on February 27, 2004.
The colored lines show various models folded through the detector
response.
The green line and the error weighted residuals are for the best fit model
listed in Table~\ref{bbody} column 3.
The blue line is the best fit model to the spectrum of February 2, 2000
(Table~\ref{bbody} column 2). The red line shows the best
fit model to the \xmm-RGS spectrum of October 27, 2003 
\citep{devries04}; it has been cut off at the wavelength limit of the
RGS instruments (38~\AA).
\label{ddt}}
\end{figure*}

\section{Observation and data analysis}

\chandra\ observed \rxj\ for 35~ks 
on February 27, 2004 (Obs. ID 5305) as part of its DDT program.
For our analysis we used the cleaned event list available from the
\chandra\ X-ray Center. Spectral extraction and
ancillary response files were made using the
standard \chandra\ reduction package CIAO v.3.0.2.
For comparison we also analyzed
the longest
archival LETGS observation, made on February 2, 2000 
\citep[Obs. ID 745, see also][]{kaplan03b}, and the six
archival \xmm\ Reflective Grating Spectrometer  (RGS) data sets discussed
in \citet{devries04}.

The LETGS DDT spectrum is shown in Fig.~\ref{ddt}.
The spectrum is clearly harder than the LETGS spectrum of February 2000.
This confirms the discovery of spectral evolution of \rxj\ \citep{devries04}.
However, it is clear that the source has continued to evolve since the
last \xmm\ observation of 27 October 2003, as the
attenuation of the emission above 23~\AA, which increased with time and
was very prominent in the last RGS spectrum,
has almost disappeared.
In fact, unlike the last RGS spectrum both LETGS spectra
can be fitted reasonably well with pure blackbody models, but with different
temperatures. The new spectrum requires a hotter blackbody temperature
of $kT_\infty \sim 100$~eV than that of February 2000, which
is best fitted with a temperature of $\sim 85$~eV. The precise values for
temperature depend on the modeling assumptions (Table~\ref{bbody}), 
e.g. whether we allow
the interstellar absorption to vary from one observation to another 
(columns \nh\ free) or whether we fit the two spectra simultaneously,
forcing the absorption parameters for the two models to be equal. 

\begin{table*}[b]
\begin{center}
\caption{Best fit blackbody models.\label{bbody}}

\begin{tabular}{lllll}
\tableline\tableline\noalign{\smallskip}
{} & \multicolumn{2}{c}{2/2/2000}  & \multicolumn{2}{c}{27/2/2004}\\
 Parameter & \nh\ free &  
 \multicolumn{2}{c}{jointly fitted \nh}  & 
 \nh\ free\\
\tableline\noalign{\smallskip}

$kT_{\infty}$\ (eV)& $82.9\pm1.5$ &$86.4\pm2.4$ &$100.6\pm1.1$& $103.4\pm1.3$\\
$R$ (km) & $5.36\pm0.12$ &$4.63\pm0.07$&$3.43\pm0.04$& $3.13\pm0.05$\\
\nh\ ($10^{20}\,{\rm cm}^{-2}$)
                  & $1.28\pm0.12$ &\multicolumn{2}{c}{$0.90\pm0.07$} 
                                                             & $0.63\pm0.09$\\
C-statistic/\#bins& 727.5/720 & 735.8/719   & 780.6/719    
        &771.3/719\\

\noalign{\smallskip}\tableline
\end{tabular}
\tablecomments{
The models were fitted using the spectral analysis program
\xspec\, \citep{xspec}.
Errors are 68\% confidence limits.
The statistic used to optimize the fit
was the C-statistic \citep{cash79}.
We assume a distance of 300~pc \citep{kaplan03b}.
}
\end{center}
\end{table*}

We note that the spectral fitting can be further improved for both LETGS
spectra by an additional soft {\em emission} component,
which contributes to $\lambda > 50$~\AA.
However, we
do not want to overemphasize this. First, it may be the result of
calibration uncertainties regarding the instrument sensitivity.
Secondly, 
assuming that the additional component
is real and caused by thermal emission,
a very cool blackbody is required ($kT_\infty \sim 23$~eV),
with an emitting area too large for a neutron star,
corresponding to a radius of $\sim 400$~km 
\citep[following][we assume here, and throughout the
rest of the text a distance of 300~pc]{kaplan03b}.

\vbox{
\centerline{
\psfig{figure=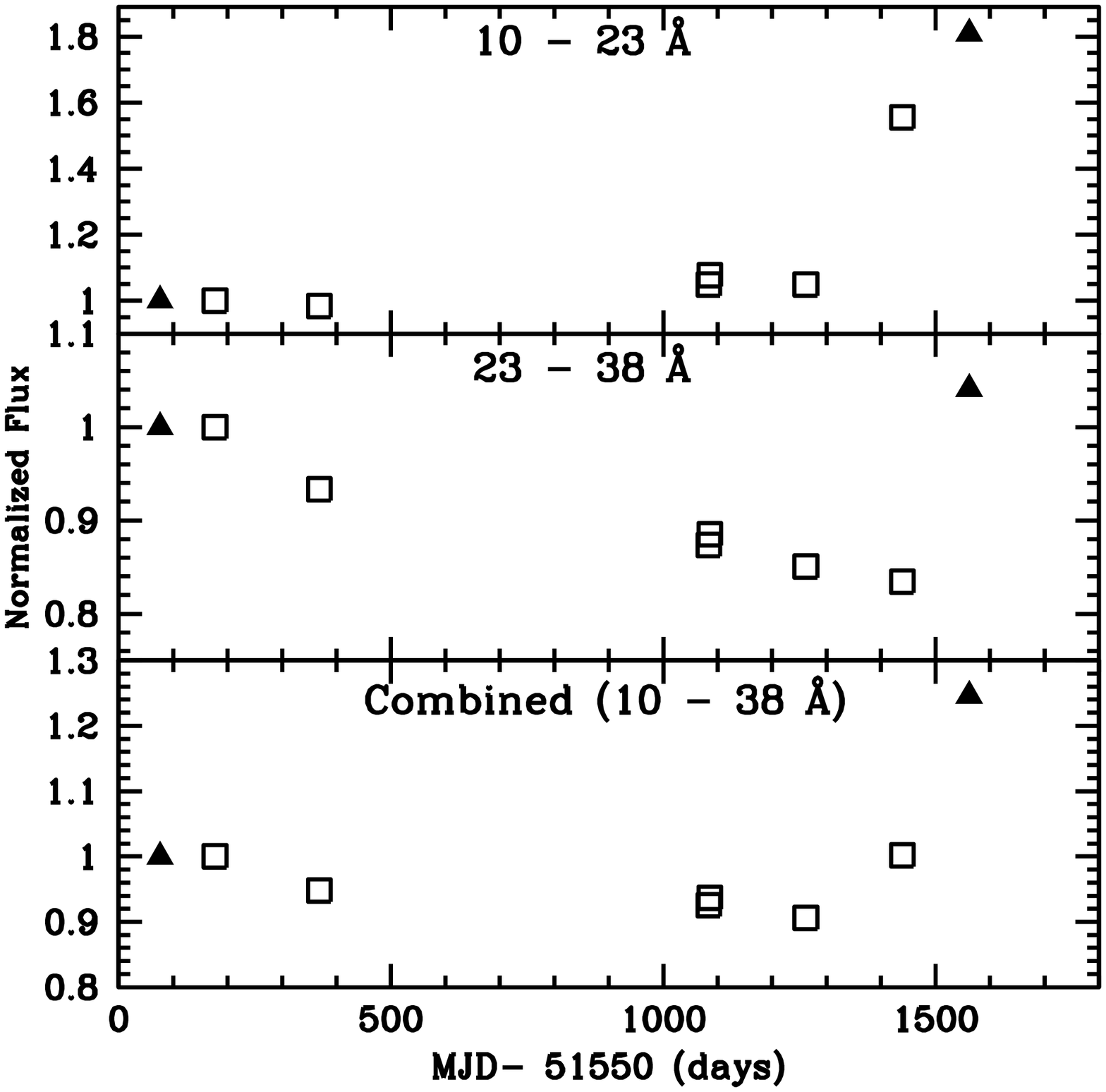,width=0.85\columnwidth}}
\figcaption{The flux evolution of \rxj\ for the wavelength ranges
10-13 \AA, 23-38 \AA\ and the combined range, determined from
 \xmm-RGS (open squares) and \chandra-LETGS spectra (triangles)
The fluxes are normalized to the fluxes of
the first RGS and LETGS observations respectively, which are thus
per definition 1.
\label{flux}}
}

The observed increase of the emission above 23~\AA\ with respect to the
last RGS observation is very interesting,
but also somewhat unfortunate,
as the broad spectral range of the LETGS would have allowed to put
better constraints on the spectral shape of the low energy attenuation.
For instance, 
\citet{devries04} 
showed that the deviations from a blackbody spectrum apparent from
the \xmm-RGS spectra,
could be either modeled by a broad Gaussian absorption component, or
by a blackbody modified by a multiplicative power law. 
\citet{devries04} 
favored the characterization by a power law times a blackbody,
as it provided good fits to the data with fewer parameters:
Apart from the blackbody parameters, only the power-law slope had to be
determined.
Although there is no clear physical
rational for such a model, the spectral evolution could be characterized
by just one parameter: the power-law slope. Changing the power-law
slope results in a simultaneous hardening and supression of long wavelength
emission, which
describes well the spectral evolution up to October, 2003,
see Fig.~\ref{flux}.
\footnote{In order to reduce the effects of 
x[systematic uncertainties in the sensitivity of the two instruments,
we have normalized the fluxes to those of the first observation 
with each instrument. The normalized fluxes can be converted to
absolute fluxes by multiplication with
$(4.3\pm0.1)\times10^{-12}$~erg\,s$^{-1}$cm$^{-2}$ (23-38~\AA)
and $(1.7\pm0.2)\times10^{-12}$~erg\,s$^{-1}$cm$^{-2}$\ (10-23~\AA).}

The alternative spectral model discussed in \citet{devries04}, 
a blackbody with
Gaussian absorption, e.g. caused by proton cyclotron absorption, has
three additional parameters: central energy, $E_0$, width, $\sigma_E$,
and normalization.
Such a model has been considered by \citet{vankerkwijk04} for RX J1605.3+3249,
and for \rxj\ by \citet{haberl03b},
who found a central energy of 271~eV and width $\sigma_E = 64$~eV. 
Reanalyzing the RGS spectra for this study, 
and simultaneously fitting all spectra with a blackbody model with
Gaussian absorption with one central energy and width gives
$E_0 = 306\pm 8$~eV (40.5~\AA) and $\sigma_E = 137\pm8$~eV. 
The central energy is outside the RGS spectral range
and the width is so large that, as far as the RGS
spectra are concerned, the Gaussian absorption component
is indistinguishable from an exponential absorption feature. However,
the RGS spectra are inconsistent with the best-fit Gaussian parameters
found by \citet{haberl03b}, mainly because the RGS spectra require 
a broader absorption feature than a Gaussian absorption with
$\sigma_E = 64$~eV.

Whatever the best model to describe the deviations from a blackbody
spectrum, the new LETGS spectrum breaks the trend that
a spectral hardening is accompanied by a decrease of long wavelength
emission.
The flux in the spectral range
from 10-90 \AA\ has slightly increased since February 2000 by 8\%,
but the best-fit parameters of the blackbody model imply
that the total blackbody flux has hardly changed.
Taken at face value this would imply that the emission area must have
decreased (Table~\ref{bbody}).

\begin{figure*}
\centerline{
  \psfig{figure=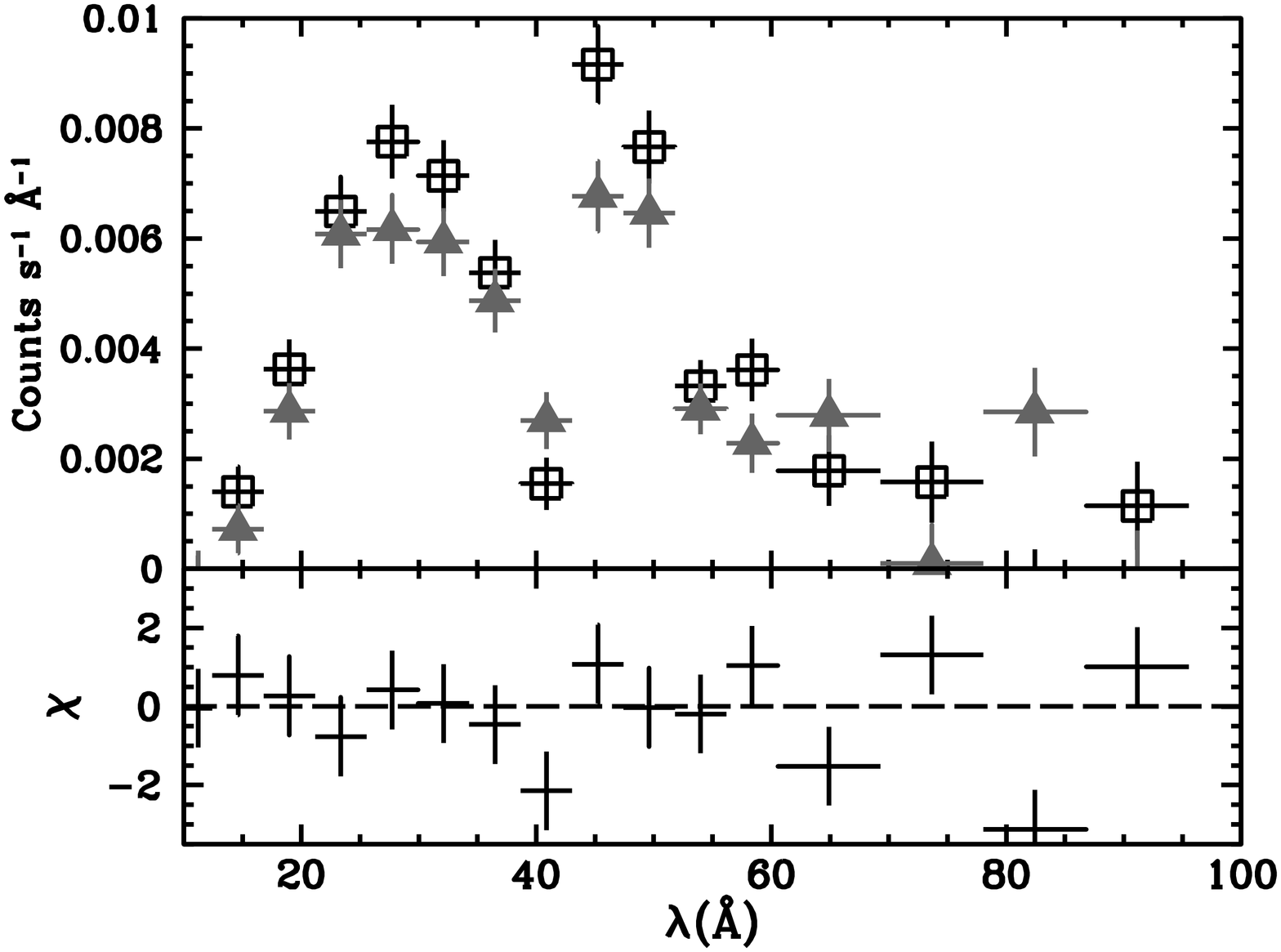,angle=-90,width=0.5\textwidth}
  \psfig{figure=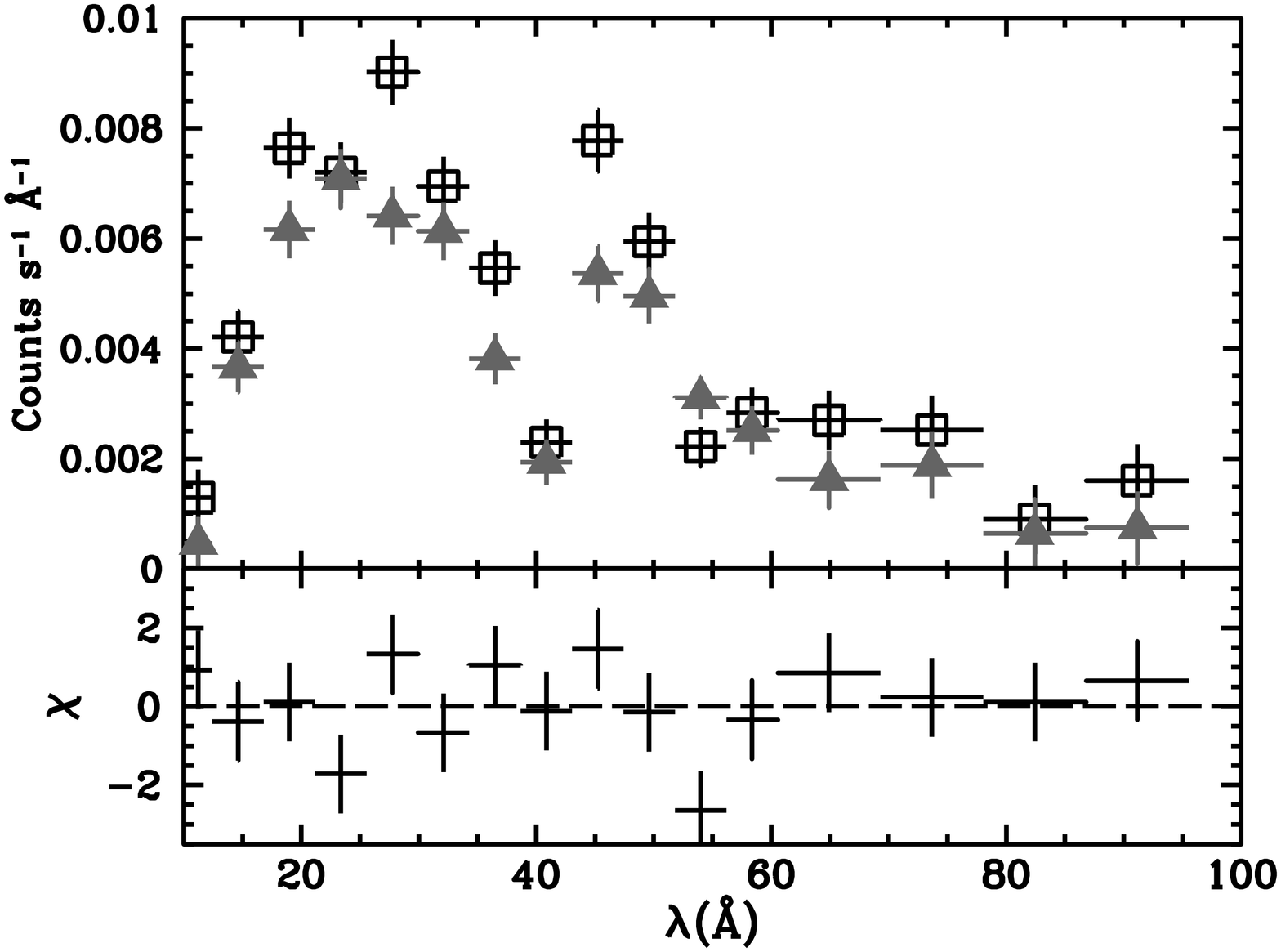,angle=-90,width=0.5\textwidth}}
\caption{
Phase resolved LETGS spectra obtained from 
the February 2000 (left) and February 2004 observations (right).
The spectral extraction was done for phases 0.125 - 0.375 (open squares) 
and 0.625-0.875 (filled triangles). The residuals take into account the
normalization differences.
\label{phases}}
\end{figure*}

It is of interest that apart from a long term change in the spectrum,
\xmm\ CCD spectroscopy indicates that the spectral shape is also a function
of pulse phase \citep{cropper01,haberl03b}, 
and is in itself also subject to evolution \citep{devries04}.
Unfortunately, the RGS timing resolution ($\sim$4~s) 
is not sufficient to
check the shape of the long wavelength attenuation as a function of pulse
phase.
On the other hand,
the timing resolution of \chandra's HRC-S detector is $\sim4$~ms.
We therefore folded the events of both the February 2000 and 2004
observation with the 8.3911~s period 
\citep[see][for the latest timing results]{cropper04}, and added a timing
offset in order to align the folded light curves 
so that the maximum emission occurs at phase 0.25 and the
minimum at 0.75. We then extracted spectra for the phase bins 0.125 - 0.375
and 0.625-0.875 (Fig.~\ref{phases}).

Apart from an obvious variation in brightness between on and off peak,
the overall shapes of the spectra do not change much.
In order to test this we statistically compared the two spectra
for each observation by direct comparison of the countrates per spectral bin.
We assumed that the spectra are the same in both phase bins, except for
a normalization factor.
The best fitting normalization factors, i.e. the ratio
in flux between on and off peak emission, are
$1.19\pm0.06$\ (Feb. 2000) and $1.22\pm0.05$ (Feb 2004),
which is consistent with previous published pulse profiles 
\citep{cropper01,devries04}.
Comparing the on peak spectra with the off peak spectra multiplied by
the normalization factor we obtained
respectively $\chi^2/dof = 23.38/15$\ and $\chi^2/dof = 17.77/15$,
which corresponds to probabilities of 8\% and 27\% that on and off peak
spectra are similar.
Our limits for variations are still consistent with the results
of \citet{cropper01}, 
who found variable absorption of the order of 
$\Delta$\nh$ = 4\times10^{19}$~cm$^{-2}$.
Note that
although the LETGS has a higher spectral resolution,
the \xmm\ CCD instruments have a much higher effective area.
Phase resolved spectroscopy with the LETGS does, however,
indicate that whatever the phase dependent spectral changes, they must be 
rather subtle and are probably caused by broad features.

\section{Discussion}
Following the discovery of long term spectral changes in the X-ray emission
of \rxj\ by \xmm, the new \chandra\ DDT observation shows that the
X-ray spectrum of \rxj\ has continued to evolve. 
In particular
the spectrum has remained hard from October 2003 to February 2004, but,
somewhat unexpected,
the attenuation of emission at wavelengths longer than 23~\AA\ 
has disappeared.

A simple minded explanation would be that part of \rxj\ surface must have 
heated up in the course of the last 4 years, initially accompanied by
absorption, but which disappeared between October 2003 and February 2004.
The absorption mechanism could be proton cyclotron absorption, as
was invoked to explain the broad absorption features in the spectra of
other isolated neutron stars.

However, such an explanation is not completely
consistent with the evolution of the spectrum of \rxj. First of all,
if the surface is heated up by a starquake \citep{larson02}, or by
increased accretion from the interstellar medium,
we would expect that an additional hot spot at the surface would appear,
or that the existing hot spot would become hotter. As a result the total
blackbody flux would increase. This is, however, in 
disagreement with our best-fit blackbody parameters and the flux
decrease between 45-55~\AA\ (Fig.~\ref{ddt}). 
Note that heating as a result of accretion from the interstellar medium 
is also unlikely given the the high proper motion 
of the neutron star \citep[$97\pm12$~mas/yr,][]{motch03}.

Our suggestion that the spectral evolution of \rxj\  is due
to precession \citep{devries04}, 
can at least qualitatively explain the results.
It assumes that the emission remains unchanged, but that we see the
hot spot under different viewing angles.
Free precession has been observed in some pulsars \citep[e.g.][]{stairs}.
Recently \citet{wasserman03} has calculated the expected precession period
caused by an oblique magnetic field and magnetic stresses in the neutron
crust. His model requires a Type II superconducting interior.
The expected precession period scales as $P_p\propto P_0B^{-1}$, with $P_0$ the
rotation period. 
For \rxj, which has an inferred magnetic field of
$B=3\times 10^{13}$~G \citep{cropper04},
the expected precession period is $P_p\sim4$~yr, which is consistent
with the timescale of the spectral evolution. 
Moreover, \xmm\ CCD spectra indicate that a modulation of the spectrum
with pulse phase is present, and the fact that pulsation
can be observed at all requires anisotropic surface emission \citep{cropper01}.
As the phase modulation of the spectrum also implies a spectral change
with viewing angle, it is not unreasonable to assume that precession gives
rise to a similar modulation of the spectrum with precession phase.

Theoretically a strong effect of the viewing angle on the observed spectrum
from a highly magnetized atmosphere is to be expected.
The reason is that protons and electrons in the neutron star atmosphere
are constrained to move along the magnetic field lines.
This results in
radiation that is strongly polarized and angle dependent.
This is further enhanced for magnetic fields $B \gtrsim 10^{13}$~G,
for which vacuum polarization becomes important. This is
a quantum electrodynamics effect that changes 
the polarization mode as the photon traverses a density gradient 
\citep{oezel01,ho04}.

The new \chandra\ observation is not able to confirm that the 
attenuation has an approximate Gaussian shape, as may be expected for
proton cyclotron absorption. However, as the spectrum of \rxj\
continues to evolve, future \chandra-LETGS observations 
may be able to constrain the spectral model further. 
Long term monitoring is necessary to test the precession idea,
since it implies a cyclic spectral evolution.
\citet{kaplan03b} suggested that \rxj\ is an off-beam radio pulsar.
If it is indeed precessing, there is a possibility that at some point
during its cycle the radio beam will be directed toward the earth.

Finally, we point out that other isolated neutron stars may show similar
behavior, indicative of precession. Further observations of those sources
is therefore important for increasing our understanding of
both the hot surface of neutron stars and for probing
their internal structure.

\acknowledgements
We thank Dr. Harvey Tananbaum for allowing
\rxj\ to be observed as part of the 
DDT program, and for his continued interest
in the results. We thank Rob van der Meer for his assistance with the LETGS
data analysis.

\end{document}